# Graphene oxide 2D films integrated with nanowires and ring resonators for enhanced nonlinear optics

David J. Moss

Optical Sciences Centre, Swinburne University of Technology, Hawthorn, VIC 3122, Australia

## ABSTRACT

We report enhanced nonlinear optics in nanowires, waveguides, and ring resonators by introducing layered two-dimensional (2D) graphene oxide (GO) films through experimental demonstration. The GO films are integrated on silicon-on-insulator nanowires (SOI), high index doped silica glass, and silicon nitride (SiN) waveguides and microring resonators (MRRs), to demonstrate an improved optical nonlinearity including Kerr nonlinearity and four-wave mixing (FWM). By using a large-area, transfer-free, layer-by-layer GO coating method with photolithography and lift-off processes, we integrate GO films on these complementary metal-oxide-semiconductor (CMOS)-compatible devices. For SOI nanowires, significant spectral broadening of optical pulses in GO-coated SOI nanowires induced by self-phase modulation (SPM) is observed, achieving a high spectral broadening factor of 4.34 for a device with a patterned film including 10 layers of GO. A significant enhancement in the nonlinear figure of merit (FOM) for silicon nanowires by a factor of 20 is also achieved, resulting in a FOM > 5. For Hydex and SiN waveguides, enhanced FWM in the GO-coated waveguides is achieved, where conversion efficiency (CE) enhancements of up to 6.9 dB and 9.1 dB relative to the uncoated waveguides. For MRRs, an increase of up to ~10.3 dB in the FWM CE is achieved due to the resonant enhancement effect. These results reveal the strong potential of GO films to improve the nonlinear optics of nanowires, waveguides, and ring resonators.

## 1. INTRODUCTION

The 3$^{rd}$ order nonlinear optical response has found wide ranging applications in telecommunications, metrology, astronomy, ultrafast optics, quantum photonics, and many other areas [1-4]. Nonlinear integrated photonic devices based on the Kerr effect ($n_2$) in particular offer far superior processing speeds compared to electronic devices as well as the added benefits of compact footprint, low power consumption, high stability, and low-cost mass production, all of which are important for high-speed signal generation and processing in optical communication systems [5-7].

While silicon-on-insulator nanowires (SOI) has shown itself to be a leading platform for integrated photonic devices, it suffers from strong two-photon absorption (TPA) at near-infrared wavelengths, which greatly limits the nonlinear performance [5, 6], and this has motivated the use of highly nonlinear materials on chips. Other complementary metal-oxide-semiconductor (CMOS) compatible platforms including high index doped silica glass [8, 9] and silicon nitride (SiN) [10, 11] have a much lower TPA, but they hamper the nonlinear performance due to a comparatively low Kerr nonlinearity. To overcome these limitations, two-dimensional (2D) layered graphene oxide (GO) has received much attention among the various 2D materials due to its ease of preparation as well as the tunability of its material properties [12-20]. Previously, we reported GO films with a giant Kerr nonlinear response about 4-5 orders of magnitude higher than that of silicon and SiN [16] and demonstrated enhanced four-wave mixing (FWM) in doped silica waveguides and microring resonators (MRRs) integrated with GO films [21, 22]. Here, we demonstrate enhanced nonlinear optics in SOI nanowires [23] and SiN waveguides [24] integrated with 2D layered GO films. Owing to the strong light-matter interaction between the integrated waveguides and the highly nonlinear GO films, self-phase modulation (SPM) measurements are performed to show significant spectral broadening enhancement for SOI nanowires coated with patterned films of GO. The dependence of GO's Kerr nonlinearity on layer number and pulse energy shows interesting physical insights and trends of the layered GO films in evolving from 2D monolayers to quasi bulk-like behavior. We obtain significant enhanced nonlinear performance for the GO hybrid devices as compared with the bare waveguides, achieving the nonlinear parameter of GO-coated SOI nanowires up by 16 times, with the nonlinear figure of merit (FOM) increasing over 20 times to FOM > 5. We obtain a significant improvement in the FWM conversion efficiency (CE) of ≈ 6.9 dB for a uniformly coated Hydex waveguide with 2 layer of GO and ≈ 9.1 dB for a patterned SiN waveguide with 5 layers of GO. For MRRs, we achieve up to ~10.3 dB in the FWM CE. These results confirm the strong potential of introducing 2D layered GO films into CMOS compatible photonic platforms to realize high-performance nonlinear photonic devices.

## 2. ENHANCED KERR NONLINEARITY IN GO-COATED SILICON-ON-INSULATOR NANOWIRES

Figure 1a shows a schematic of an SOI nanowire waveguide integrated with a GO film. The fabrication of the SOI nanowire can be achieved via either deep ultraviolet photolithography or e-beam lithography followed by inductively coupled plasma etching, all of which are mature silicon device fabrication technologies [25, 26]. The GO film coating, with a thickness of ~2 nm per layer [23], can be achieved using solution-based methods that yield layer-by-layer film deposition [20, 21, 27]. As compared with the sophisticated transfer processes for other 2D materials such as graphene and TMDCs [28, 29], these coating methods enable transfer-free and conformal film coating, with high fabrication stability, repeatability, precise control of the film thickness (i.e., number of layers), and extremely good film attachment onto integrated photonic devices [30]. Figure 1b shows a microscope image of a fabricated SOI chip with a 0.4-mm-long opened window. Apart from allowing precise control of the placement and coating length of the GO films that are in contact with the SOI nanowires, the opened windows also enabled us to test the performance of devices having a shorter length of GO film but with higher film thicknesses (up to 20 layers). This provided more flexibility to optimize the device performance with respect to SPM spectral broadening. Figure 1c shows the scanning electron microscopy (SEM) image of an SOI nanowire conformally coated with 1 layer of GO. Note that the conformal coating (with the GO film coated on both the top surface and sidewalls of the nanowire) is slightly different to earlier work where we reported doped silica devices with GO films only coated on the top surface of the waveguides [30, 31]. As compared with doped silica waveguides, the SOI nanowires allow much stronger light-material interaction between the evanescent field leaking from the waveguide and the GO film, which is critical to enhance nonlinear optical processes such as SPM and FWM. Figure 1d shows the successful integration of GO films which is confirmed by the representative D (1345 cm-1) and G (1590 cm-1) peaks of GO observed in the Raman spectrum of an SOI chip coated with 5 layers of GO. Microscope images of the same SOI chip before and after GO coating are shown in the insets, which illustrate good morphology of the films.

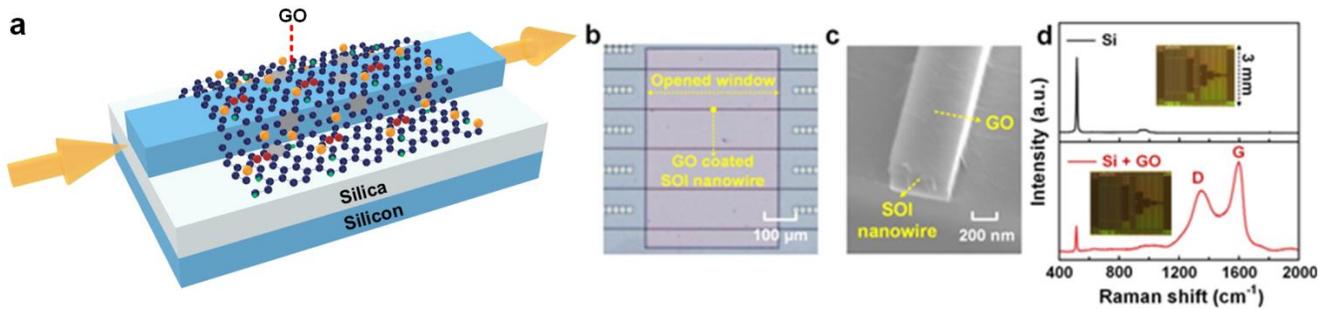

Figure 1. (a) Schematic illustration of a GO-coated SOI nanowire waveguide. (b) Microscope image of a fabricated SOI chip with a 0.4-mm-long opened window. (c) Scanning electron microscopy (SEM) image of a SOI nanowire conformally coated with 1 layer of GO. (d) Raman spectra of an SOI chip without GO and with 5 layers of GO. Insets show the corresponding microscope images.

Figure 2 shows the results of the SPM experiments. Figure 2a-i shows the normalized spectra of the optical pulses before and after transmission through the SOI nanowires with 2.2-mm-long, 1−3 layers of GO, together with the output optical spectrum for the bare SOI nanowire, all taken with the same pulse energy of ~51.5 pJ (i.e., ~13.2 W peak power, excluding coupling loss) coupled into the SOI nanowires. As compared with the input pulse spectrum, the output spectrum after transmission through the bare SOI nanowire exhibited measurable spectral broadening. This is expected and can be attributed to the high Kerr nonlinearity of silicon. The GO-coated SOI nanowires, on the other hand, show much more significantly broadened spectra as compared with the bare SOI nanowire, clearly reflecting the improved Kerr nonlinearity of the hybrid waveguides. Figure 2a-ii shows the corresponding results for the SOI nanowires with 0.4-mm-long, 5−20 layers of GO, taken with the same coupled pulse energy as in Figure 2a-i. The SOI nanowires with a shorter GO coating length but higher film thicknesses also clearly show more significant spectral broadening as compared with the bare SOI nanowire. We also note that in either Figure 2a-i or 2a-ii, the maximum spectral broadening is achieved for a device with an intermediate number of GO layers (i.e., 2 and 10 layers of GO in a-i and a-ii, respectively). This could reflect the trade-off between the Kerr nonlinearity enhancement (which dominates for the device with a relatively short GO coating length) and loss increase (which dominates for the device with a relatively long GO coating length) for the SOI nanowires with different numbers of GO layers.

Figures 2b-i and b-ii show the power-dependent output spectra after going through the SOI nanowires with (i) 2 layers and (ii) 10 layers of GO films. We measured the output spectra at 10 different coupled pulse energies ranging from ~8.2 pJ to ~51.5 pJ (i.e., coupled peak power from ~2.1 W to ~13.2 W). As the coupled pulse energy was increased, the output spectra showed increasing spectral broadening as expected. We also note that the broadened spectra exhibited a slight asymmetry. This was a combined result of both the asymmetry of the input pulse spectrum and the free-carrier effect of silicon including both the free carrier absorption (FCA) and free carrier dispersion (FCD). Since the time response for the generation of free carries is slower compared to the pulse width, there is a delayed impact of FCA on the pulse shape, which leads the spectral asymmetry of the optical pulses. The FCD further broadens the asymmetry induced by FCA, resulting in more obvious spectral asymmetry at the output.

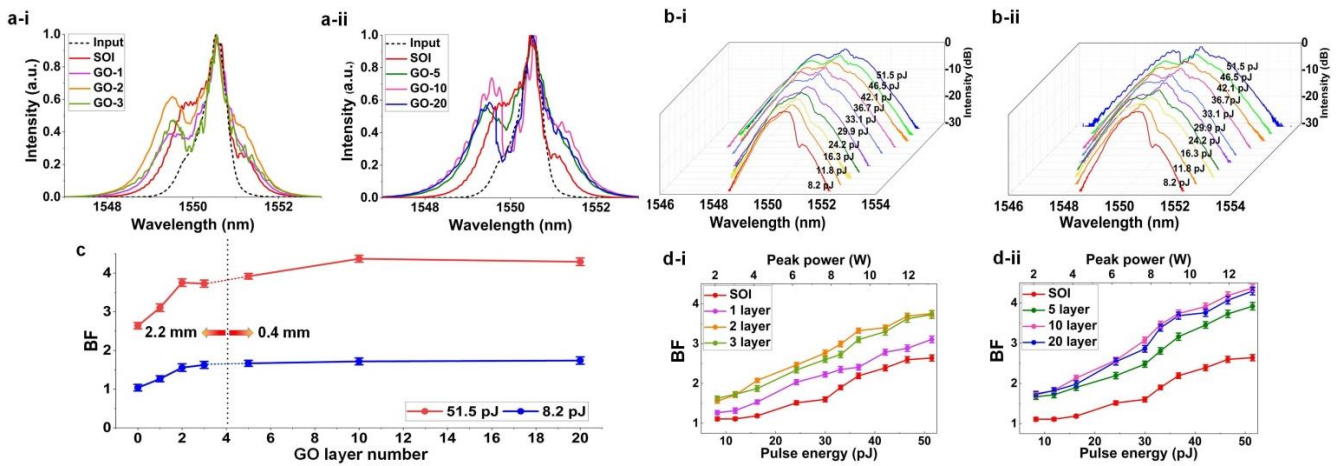

Figure 2. SPM experimental results. (a) Normalized spectra of optical pulses before and after going through the GO-coated SOI nanowires at a coupled pulse energy of ~51.5 pJ. (b) Optical spectra measured at different pulse energies for the GO-coated SOI nanowires. (c) BFs of the measured output spectra versus GO layer number at fixed coupled pulse energies of 8.2 pJ and 51.5 pJ. (d) BFs of the measured output spectra versus coupled pulse energy (or coupled peak power). In (a), (b) and (d), (i) and (ii) show the results for the SOI nanowires with 2.2-mm-long, 1−3 layers of GO and with 0.4-mm-long, 5−20 layers of GO, respectively. In (a), (c) and (d), the corresponding results for the bare SOI nanowires are also shown for comparison.

To quantitatively analyze the spectral broadening of the output spectra, we introduce the concept of a broadening factor (BF, defined as the square of the pulse'rms spectral width at the waveguide output facet divided by the corresponding value at the input [31]). Figure 2c shows the BFs of the measured output spectra after transmission through the bare SOI nanowire and the GO-coated SOI nanowires at pulse energies of 8.2 pJ and 51.5 pJ. The GO-coated SOI nanowires show higher BFs than the bare SOI nanowires (i.e., GO layer number = 0), and the BFs at a coupled pulse energy of 51.5 pJ are higher than those at 8.2 pJ, agreeing with the results in Figures 2a and 2b, respectively. At 51.5 pJ, BFs of up to 3.75 and 4.34 are achieved for the SOI nanowires with 2 and 10 layers of GO, respectively. This also agrees with the results in Figure 2a − with the maximum spectral broadening being achieved for an intermediate number of GO layers due to the trade-off between the Kerr nonlinearity enhancement and increase in loss. The BFs of the output spectra versus coupled pulse energy are shown in Figures 2d-i and 2d-ii for the SOI nanowires with 1−3 layers and 5−20 layers of GO, respectively. The BFs increase with coupled pulse energy, reflecting a more significant spectral broadening that agrees with the results in Figure 2b.

## 3. ENHANCED FWM IN GO-COATED HYDEX AND SIN WAVEGUIDES

Figure 3a shows the GO-coated integrated waveguides made from high-index doped silica glass [5], with a cross section of 2 μm × 1.5 μm. The integrated waveguide is surrounded by silica except that the upper cladding is removed to enable coating the waveguide with GO films. The GO films, with a thickness of about 2 nm per layer, were introduced on the top of the integrated waveguide in order to introduce light-material interaction with the evanescent field leaking from the integrated waveguide. The Kerr coefficient of GO is on the order of $10^{-15} \sim 10^{-14}$ m$^2$/W [16, 19], which is slightly lower than that of graphene (~$10^{-13}$ m$^2$/W) [19, 32, 33], but still orders of magnitude higher than that of high-index doped silica glass (~$10^{-19}$ m$^2$/W) and silica (~$10^{-20}$ m$^2$/W) [5]. The waveguides were fabricated via CMOS compatible processes [34, 35]. First, high-index doped silica glass films ($n$ = ~1.60 at 1550 nm) were deposited using standard plasma enhanced chemical vapour deposition (PECVD), then patterned using deep UV photo-lithography and etched via reactive ion etching (RIE) to form waveguides with exceptionally low surface roughness. After that, silica glass ($n$ = ~1.44 at 1550 nm) was deposited via PECVD and the upper cladding of the integrated waveguides was removed by chemical mechanical polishing (CMP). Finally, the GO film was coated on the top surface of the chip by a solution-based method that yields layer-by-layer deposition of GO films, as reported previously [20]. An image of the integrated waveguide incorporating two layers of GO is shown in Figure 3b, which illustrates that the morphology is good, leading to a high transmittance of the GO film on top of the integrated waveguide. The integration of GO onto the waveguide is confirmed by Raman spectroscopic measurements (Figure 3c) that show the representative D (1345 cm$^{-1}$) and G (1590 cm$^{-1}$) peaks of GO [16].

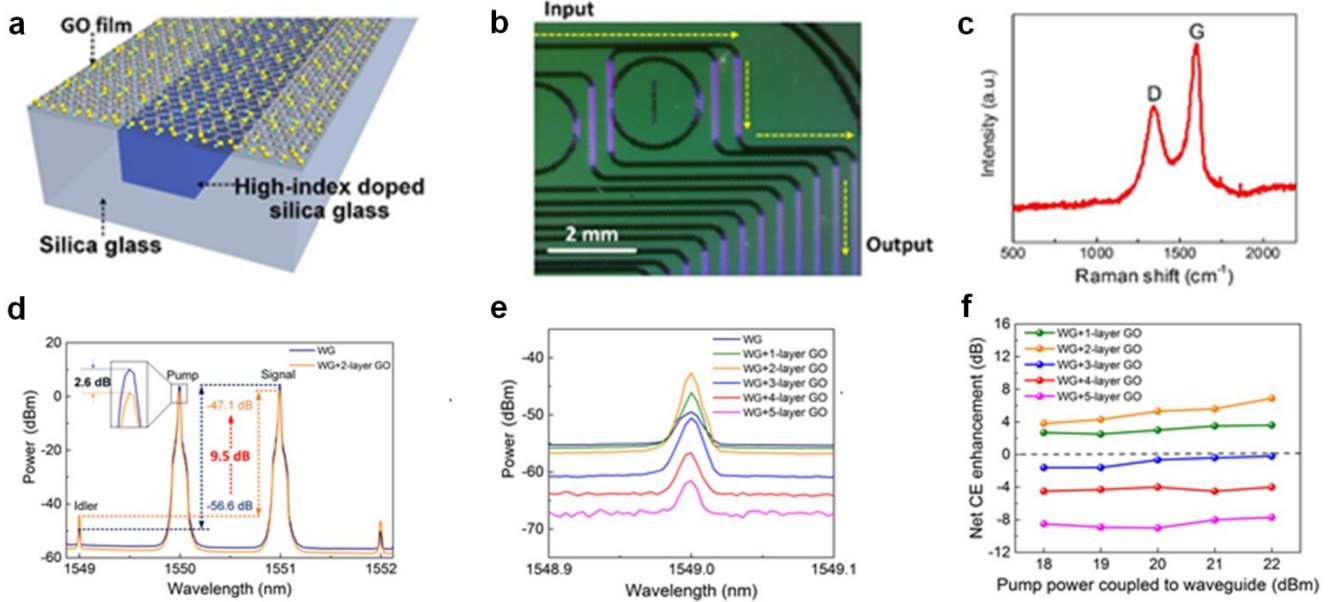

Figure 3. (a) Schematic illustration of hybrid Hydex waveguides integrated with GO. (b) Image of the hybrid Hydex integrated waveguide with GO. (c) Raman spectra of GO on the integrated chip. (d) FWM spectra of the integrated waveguide without GO and with 2 layers of GO. (e) Zoom in spectra of the generated idlers after FWM in the waveguide with 0 to 5 layers of GO. (f) Net CE enhancements for various pump powers coupled to the waveguide with 1 to 5 layers of GO.

The FWM spectra of a 1.5-cm-long integrated waveguide without GO and with 2 layers of GO are shown in Figure 3d. For comparison, we kept the same pump power of ~30 dBm before the input of the waveguide, which corresponded to ~22 dBm pump power coupled into the waveguide. It can be seen that although the hybrid integrated waveguide had additional propagation loss (~2.6 dB), it clearly shows enhanced idler output powers as compared with the same waveguide without GO. The CE (defined as the ratio of the output power of the idler to the output power of the signal, i.e., $P_{out, idler}/P_{out, signal}$) of the integrated waveguide with and without GO were ~-47.1 dB and ~-56.6 dB, respectively, corresponding to a CE enhancement of ~9.5 dB for the hybrid integrated waveguide. After excluding the addtional propagation loss, the net CE enhancement (defined as the improvement of the output power of the idler for the same pump power coupled to the waveguide) is 6.9 dB.

For the integrated waveguide coated with 1 to 5 layers of GO, zoom-in spectra of the generated idlers for the same pump power coupled to the waveguide (~22 dBm) are shown in Figure 3e. For the integrated waveguide coated with 1 and 2 layers of GO, there were positive net CE enhancements. When the number of GO layers was over 2, the net change in CE was negative. This is mainly due to the super-linear increase in propagation loss for increased numbers of GO layers as noted above. The output powers of the idler for various pump powers coupled to the waveguide without GO and with 2 layers of GO are shown in Figure 3f.

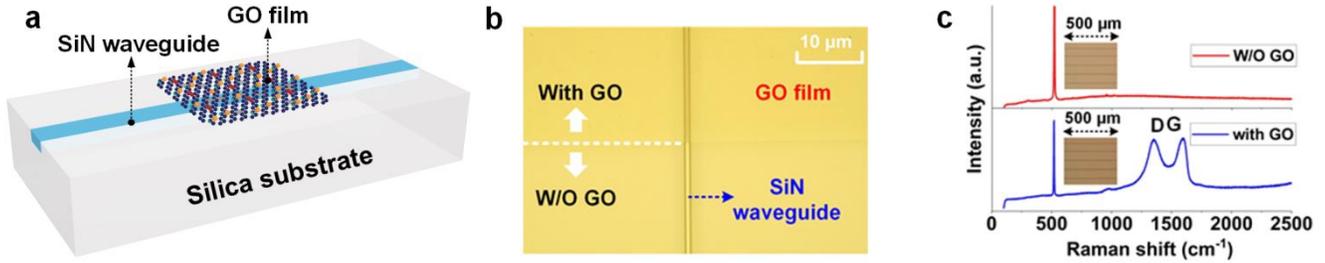

Figure 4. (a) Schematic illustration of GO-coated SiN waveguide. (b) Microscope image of a SiN waveguide patterned with 10 layers of GO. (c) Raman spectra of a SiN chip without GO and with 10 layers of GO. Insets show the corresponding microscope images.

Figure 4a shows the SiN waveguide integrated with a GO film. SiN waveguides with a cross section of 1.6 μm × 0.66 μm were fabricated via annealing-free and crack-free processes that are compatible with CMOS fabrication [36, 37]. First, a SiN layer was deposited via low-pressure chemical vapor deposition (LPCVD) in two steps, with a 370-nm-thick layer for each, so as to control strain and to prevent cracks. In order to produce high-quality films, a tailored ultra-low deposition rate (< 2 nm/ min) was used. Waveguides were then formed via a combination of deep ultraviolet lithography and fluorine-based dry etching that yielded exceptionally low surface roughness. Next, a 3-μm thick silica upper cladding layer was deposited via high-density plasma-enhanced chemical vapor deposition (HDP-PECVD) to avoid void formation. To enable the interaction between the GO films and the evanescent field leaking from the SiN waveguides, the silica upper cladding was removed using a perfectly selective CMP that left the top surface of the SiN waveguides exposed in air, with no SiN consumption and no remaining topography. Figure 4b shows a microscope image of a SiN waveguide patterned with 10 layers of GO, which illustrates the high transmittance and good morphology of the GO films. Figure 4b presents a scanning electron microscopy (SEM) image of a GO film with up to 5 layers of GO monolayers, clearly showing the layered film structure. Figure 4c shows the measured Raman spectra of a SiN chip without GO and with 10 layers of uniformly coated GO films. The successful integration of GO films is confirmed by the presence of the representative D (1345 cm-1) and G (1590 cm-1) peaks of GO.

Figure 5 shows the experimental FWM optical spectra for the SiN waveguides uniformly coated with 1 and 2 layers of GO (Figure 5a-i) together with the FWM spectrum of the bare SiN waveguide. For comparison, we kept the same power of 23 dBm for both the pump and signal before the input of the waveguides, which corresponded to 18 dBm power for each coupled into the waveguides. The difference among the baselines of the spectra reflects the difference in waveguide propagation loss for different samples. It can be seen that although the hybrid waveguide with 1 layer of GO film had an additional propagation loss of ≈7.1 dB, it clearly shows enhanced idler output powers as compared with the bare SiN waveguide. The CE of the SiN waveguides without GO and with 1 layer of GO were ≈-65.7 dB and ≈-58.4 dB, respectively, corresponding to a CE enhancement of ≈7.3 dB for the hybrid waveguide. In contrast to the positive CE enhancement for the hybrid waveguide with 1 layer of GO, the change in CE for the hybrid waveguide with 2 layers of GO was negative.

Figure 5a-ii shows the FWM spectra of the SiN waveguides with 5 and 10 layers of patterned GO films. The coupled CW pump and signal power (18 dBm for each) was the same as that in Figure 5a-i. The SiN waveguides with patterned GO films also had an additional insertion loss as compared with the bare waveguide, while the results for both 5 and 10 GO layers show enhanced idler output powers. In particular, there is a maximum CE enhancement of ≈ 9.1 dB for the SiN waveguide patterned with 5 layers of GO, which is even higher than that for the uniformly coated waveguide with 1 layer of GO. This reflects the trade-off between FWM enhancement (which dominates for the patterned devices with a short GO coating length) and loss (which dominates for the uniformly coated waveguides with a much longer GO coating length) in the GO-coated SiN waveguides. Figure 5b shows the measured CE versus pump power for the uniformly coated and patterned devices,

respectively. The plots show the average of three measurements on the same samples and the error bars reflect the variations, showing that the measured CE is repeatable. As the pump power was increased, the measured CE increased linearly with no obvious saturation for the bare SiN waveguide and all the hybrid waveguides, indicating the low TPA of both the SiN waveguides and the GO films. For the bare waveguide, the dependence of CE versus pump power shows a nearly linear relationship, with a slope rate of about 2 for the curve as expected from classical FWM theory [38]. For the GO-coated waveguides, the measured CE curves have shown slight deviations from the linear relationship with a slope rate of 2, particularly at high light powers. Figure 5c compares the CE of the hybrid waveguides with four different numbers of GO layers (i.e., 1, 2, 5, 10), where we see that the hybrid waveguide with an intermediate number of GO layers has the maximum CE. This reflects the trade-off between $\gamma$ and loss in the hybrid waveguides, which both increase with GO layer number.

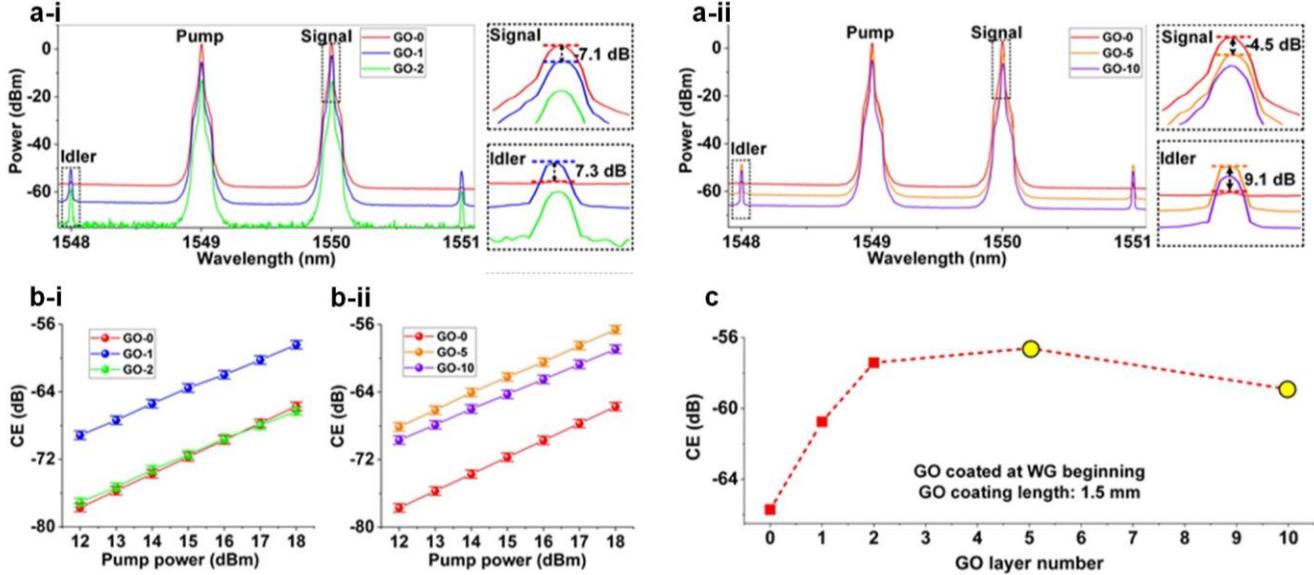

Figure 5. FWM experimental results. (a) FWM optical spectra. Insets show zoom-in view around the signal and idler. (b) CE versus pump power coupled into the waveguides. In (a) and (b), (i) shows the results for SiN waveguides uniformly coated with 1 and 2 layers of GO and (ii) shows the results for SiN waveguides patterned with 5 and 10 layers of GO. (c) Calculated CE as functions of GO layer number. In (c), the corresponding results for the bare SiN waveguide (GO-0) are also shown for comparison. The coating length is 1.5 mm and the GO coating position is at waveguide beginning.

## 4. ENHANCED FWM IN GO-COATED MICRO-RING RESONATORS

Figure 6a shows a schematic of an integrated MRR incorporating a GO film. The MRR was fabricated on a high index doped silica glass platform using CMOS compatible fabrication processes [8, 39] with CMP used as the last step to remove the upper cladding, so as to enable GO film coating on the top surface of the MRR. Benefiting from extraordinarily low linear and nonlinear loss, high index doped silica glass has been a successful integrated platform for nonlinear photonic devices [9, 40, 41]. The Kerr coefficient $n_2$ of the high index doped silica glass (~$1.3 \times 10^{-19}$ m$^2$/W) is lower than that of silicon (~$4.5 \times 10^{-18}$ m$^2$/W), while its negligible nonlinear loss even up to extremely high light intensities yields a nonlinear figure of merit (>>1) that is much higher than that of silicon (~0.3) [5]. Figure 6b shows microscopic images of an integrated MRR patterned with 50 layers of GO (~50 μm pattern length). Note that only the center ring of the 9 concentric rings (see inset) was coupled to through/drop bus waveguides to form an MRR − the rest were to aid in visual identification. Figure 6c shows the measured Raman spectra of an integrated chip (including doped silica MRRs) without GO and with 2 layers of uniformly coated GO film. The presence of the representative D (1345 cm-1) and G (1590 cm-1) peaks of GO confirms the integration of GO film onto the top surface.

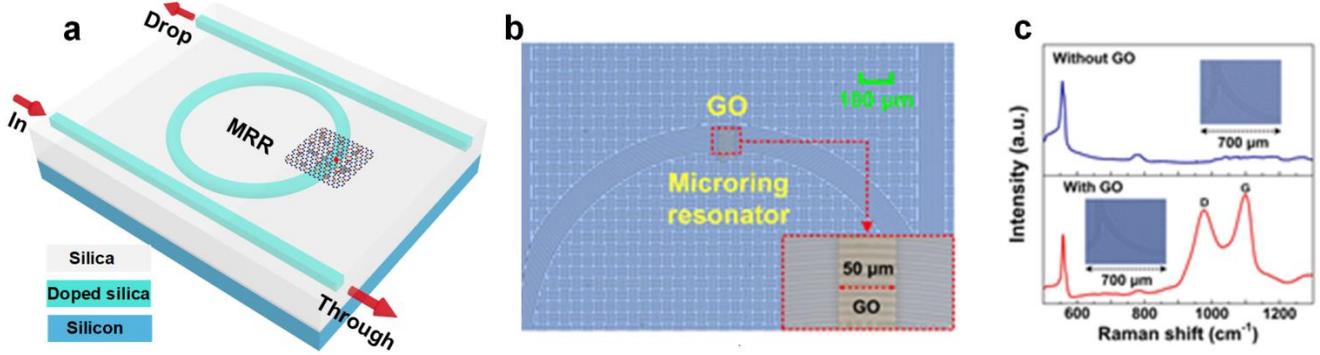

Figure 6. (a) Schematic illustration of GO-coated integrated MRR. Inset shows schematic atomic structure of GO. (b) Microscopic image of an integrated MRR patterned with 50 layers of GO. Inset shows zoom-in view of the patterned GO film. (c) Raman spectra of an integrated chip without GO and with 2 layers of GO. Insets show the corresponding microscope images.

Figure 7a shows the FWM spectra of the MRRs uniformly coated with 1−5 layers of GO, together with the FWM spectrum of the uncoated MRR. For comparison, we kept the same pump power of ~22 dBm coupled into the MRRs after excluding the mode coupling loss between the SMF array and the input bus waveguide as well as the GO-induced propagation loss of the input bus waveguide. The pump and signal had the same power and were separated in wavelength by 2 FSRs of the MRRs. As compared with the uncoated MRR, the GO-coated MRRs had an additional insertion loss (defined as the excess insertion loss of the GO-coated MRRs over the uncoated MRR), while the MRRs with 1 and 2 layers of GO clearly show enhanced idler output powers. The CE of the MRR without GO and with 1 layer of GO were ~-48.4 dB and ~-40.8 dB, respectively, corresponding to a CE enhancement of 7.6 dB for the GO-coated MRR. Figure 7b shows the FWM spectra of the MRRs with 10−50 layers of patterned GO. The GO coating length was ~50 μm and the input pump power (22 dBm) was the same as that in Figure 7a. The MRRs with patterned GO films also had an additional insertion loss as compared with the uncoated MRR, and the results for all the tested GO layer numbers show enhanced idler output powers. In particular, there is a maximum CE enhancement of ~10.3 dB for the MRR patterned with 50 layers of GO. The CE enhancement and additional insertion loss extracted from Figures 7a and b are shown in Figures 7c and d, respectively. As can be seen, the additional insertion loss increases with the GO layer number. For the MRRs with uniformly coated GO, the CE enhancement decreases with the GO layer number, whereas the MRRs with patterned GO shows the opposite trend. This could reflect the trade-off between FWM enhancement (which dominates for the patterned MRRs with a short GO coating length) and loss (which dominates for the uniformly coated MRRs with a much longer GO coating length) in the GO-coated MRRs.

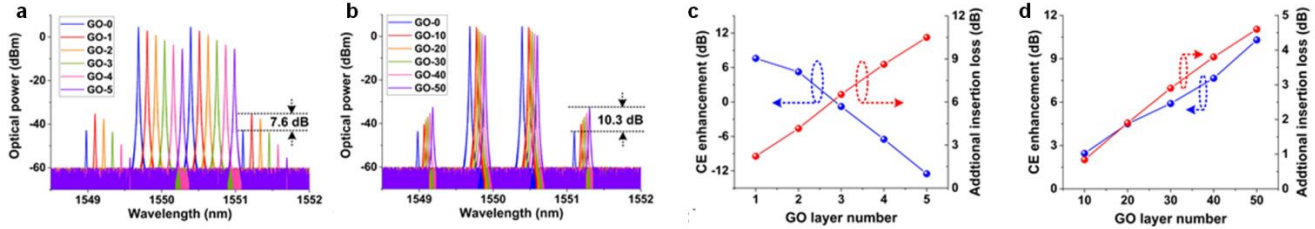

Figure 7. (a)−(b) Optical spectra of FWM at a pump power of 22 dBm for the MRRs with 1−5 layers of uniformly coated and 10−50 layers of patterned GO films, respectively. (c)−(d) CE enhancement and additional insertion loss extracted from (a) and (b), respectively. In (a) and (b), the results for uncoated MRR (GO-0) are also shown for comparison.

## 5. CONCLUSION

We demonstrate enhanced nonlinear optics including Kerr nonlinearity and FWM in nanowires, waveguides, and ring resonators incorporated with layered GO films. We achieve precise control of the placement, thickness, and length of the GO films using layer-by-layer coating of GO films followed by photolithography and lift-off. Owing to the strong mode overlap between the platforms and the highly nonlinear GO films, we achieve a high nonlinear parameter of GO coated SOI nanowires up to 16 times and an improved nonlinear FOM of up to a factor of 20. Also, we obtain a significant improvement in the FWM CE of ≈6.1 dB for a uniformly coated Hydex waveguide with 2 layer of GO and ≈9.1 dB for a patterned SiN waveguide with 5 layers of GO. For MRRs, we achieve up to ~10.3 dB in the FWM CE. Further recent theoretical work has shown that these results can be significantly improved, [42-47] aided by sophisticated integration fabrication methods. [48] This work will potentially have significant implications for on-chip microcombs and their applications. [49-112] In summary, our results verify the enhanced nonlinear optical performance of nanowires, waveguides, and ring resonators achievable by incorporating 2D layered GO films.